\documentclass[preprint,12pt]{elsarticle}

\usepackage{amssymb}
\usepackage{amsmath}
\usepackage{hyperref}
\hypersetup{
	colorlinks=true,
	linkcolor=blue,
	filecolor=magenta, 
	urlcolor=cyan,
}
\usepackage{multirow}
\usepackage[table]{xcolor}
\usepackage{hhline}

\usepackage{algorithm}
\usepackage[noend]{algpseudocode}

\journal{Computers \& Fluids}
\begin{document}

\begin{frontmatter}


 \title{Two-phase hydrodynamic model of active colloid motion}
 \author[label1]{A. Kiverin \corref{cor1}}   \ead{alexeykiverin@gmail.com}
 \author[label1]{S. Luguev}
 \author[label1]{I. Yakovenko}

 \address[label1]{Joint Institute for High Temperatures of RAS, Russia}

 \cortext[cor1]{Corresponding author. Joint Institute for High Temperatures, Izhorskaya st. 13 Bd.2, Moscow, 125412, Russia. Tel.: +7 4954844433}

\begin{abstract}
The paper presents a two-phase hydrodynamic model for the numerical simulation of collective motion in a thin layer of active colloids containing spherical microswimmers. The model accounts for three fundamental mechanisms governing the dynamics of the active colloid: the random motion of the microswimmers, their mutual collisions, and their interaction with the surrounding fluid phase. The accurate resolution of the characteristic time scales associated with each mechanism is crucial for reproducing the different dynamic modes. The model reproduces two primary modes of motion: Brownian and collective, as well as the transition between them. It is demonstrated that hydrodynamic interactions begin to play a significant role when the microswimmer velocity exceeds a critical threshold. At this point, the kinetic energy transferred to the fluid phase is sufficient to generate a noticeable feedback effect on the swimmers' motion. Conversely, a further increase in microswimmers' velocity enhances the role of collisions, causing the system to revert from a collective mode back to a Brownian-like state. A similar transition occurs at higher volume fractions of microswimmers within the colloid. 
\end{abstract}


\end{frontmatter}


\section{Introduction}
\label{sec1}
There is significant research interest in dynamic processes, evolution, and self-organization within active colloidal systems \cite{bechinger2016active}. These systems comprise active nanoparticles or microparticles capable of converting their internal energy or energy absorbed from the environment into directed motion. These particles are commonly referred to as ``motors'' or ``swimmers'', with the terms ``micromotors'' or ``microswimmers'' denoting those of the micron size. A particularly intriguing feature of active colloids is the ability of the particles to exhibit collective behavior, leading to self-organization and the emergence of directed flows. This property suggests potential applications for active particles as agents for substance delivery, including drug delivery in therapeutic contexts \cite{Lin2021}. There is also considerable interest in synthesizing complex active particles that mimic living cells, which can be used to model biological processes \cite{YangSmart}. At the same time, the evolution of biological processes themselves with account of peculiarities of microorganisms motion is also of certain interest \cite{SCHERR2015274}. Furthermore, studying the evolution of these intrinsically non-equilibrium systems can address fundamental questions in non-equilibrium statistical physics.

Among the variety of active colloidal systems, from colloidal plasmas \cite{vasiliev2023} to living biosystems \cite{dunkel2013fluid}, we focus on active suspensions \cite{senoshenko2024} and emulsions \cite{kichatov2025cluster}. These systems consist of active micro-scale solid particles or droplets dispersed in a liquid medium. Their activity can be driven by various mechanisms, including chemical reactions \cite{kichatov2025controlling}, applied magnetic fields \cite{kichatov2023pattern}, radiation absorption by nanoscale inclusions \cite{kononov2024active}, among others. A unifying characteristic of active suspensions and emulsions is the hydrodynamic interaction between the active particles and the surrounding fluid. Each particle entrains the fluid as it moves, and the resulting flow field, in turn, influences the motion of other particles, altering their trajectories. This mutual coupling can lead to a positive feedback loop, enabling the self-organization of particles' motion. This phenomenon is particularly relevant for understanding the navigation of microswimmers in established flows, ranging from plane Poiseuille flow \cite{zeng2025dispersion} to turbulent flows \cite{gupta2025can}, as well as flows in pipes \cite{das2025thermally} and blood capillaries \cite{amoudruz2025optimal}.

Two primary modeling approaches are typically employed to analyze the interaction mechanism between active particles and the surrounding fluid. The first approach is based on the dynamics of individual particles, where equations of motion include source terms that model the interaction of each particle with its environment \cite{shaebani2020computational}. The second approach uses a single-phase hydrodynamic approximation, introducing a kinetic energy source on subgrid scales through a viscous stress model \cite{slomka2015generalized, linkmann2019phase}. A common limitation of these methods is the need to incorporate specific, often ad hoc, source terms into single-phase formulations. An alternative and promising strategy is the direct numerical simulation of a two-phase flow, where the interphase interactions are modeled using established frameworks, such as the Stokes model. In this context, the main objective of the present work is to develop and investigate such a two-phase approach, analyzing the solutions it yields for a thin layer of an active suspension or emulsion.

The structure of the paper is as follows. First, the mathematical model for a two-phase active colloid system is presented, along with the numerical method employed. Then, a simple problem setup is formulated, and the calculation results are analyzed. Three basic modes of active colloid dynamics are identified and described.

\section{Mathematical model}
\label{sec2}
We consider an active system composed of spherical particles with specified size and mass, suspended in a liquid of known density and viscosity. The dynamics of this system are described using a two-phase hydrodynamic framework. This approach separately tracks the motion of each individual particle and models the surrounding liquid within the continuum mechanics approximation. The proposed two-phase description of the active system enables the resolution of all fundamental mechanisms governing the colloid's motion: the autonomous motion of each particle (microswimmer), their mutual interactions through collisions, and their hydrodynamic interaction with the surrounding liquid. The liquid is entrained by the moving particles and, in turn, provides momentum feedback that organizes their collective motion.

\subsection{Governing equations}
\label{s_sec21}
Let us first introduce the mathematical model for the motion of each particle (microswimmer). Within the proposed framework, each particle is continuously subjected to an active force that represents the energy input driving its motion. The active force has a constant magnitude, but its direction changes continuously due to stochastic noise and rotational diffusion \cite{lisin2025active}. Additionally, each particle interacts with the surrounding liquid via the Stokes drag force. Collisions between particles are modeled as perfectly elastic. The equation of motion for each active particle is given as follows.

\begin{equation}
\frac{{d{{\vec u}_p}}}{{dt}} =  {\vec F_{p}} -\frac{{\vec u_p}-{\vec u}}{\tau_{St}} 
\label{eq:1}
\end{equation}

\begin{equation}
\frac{{d{{\vec x}_p}}}{{dt}} = {\vec u_p}
\label{eq:2}
\end{equation}

\noindent where $\vec{u}_p$ is the velocity of an active particle, $\vec{x}_p$ is its position at a given time $t$, $\vec{u}$ is the velocity of the surrounding liquid, $\vec{F_{p}} = \vec{F}_{\text{active}} + \vec{F}_{\text{col}}$ is the resultant force acting on the particle, $\vec{F}_{\text{active}}$ is the active force, $\vec{F}_{\text{col}}$ is the sum of forces from elastic collisions with neighboring particles, and $\tau_{St}$ is the Stokes time, which characterizes the time scale of momentum exchange between the particle and the liquid. The Stokes time is defined as:

\begin{equation}
\tau_{St}=\frac{m_p}{3\pi\mu d_p}
\label{eq:3}
\end{equation}

\noindent here $m_p$ and $d_p$ are the mass and diameter of the particle, and $\mu$ is the dynamic viscosity of the surrounding liquid.

The liquid phase of the colloidal suspension is assumed to be incompressible and viscous. The governing equations for the liquid motion are given by:

\begin{equation}
   \nabla \cdot {\vec u} = 0  
  \label{eq:4}
\end{equation}

\begin{equation}
     \frac{{\partial {\vec u} }}{{\partial t}} + \left({\vec u}\cdot\nabla\right){\vec u}  = - \frac{\nabla p}{{\rho}} + \frac{\mu}{\rho}\nabla^2{\vec u} - \frac{1}{\rho~V}\sum\limits_{N_p}m_p \frac{d{\vec u_p}}{dt} 
  \label{eq:5}
\end{equation}

\noindent where $p$ is the pressure, $\rho$ is the density, and $V$ is the volume of fluid containing $N_p$ particles. The volume fraction of particles in the suspension, $n_p$, is related to $N_p$ by $N_pV_p=n_pV$, where $V_p$ is the volume of a single particle. Note, that in the framework of the accepted mathematical model we assume that all the particles are much smaller than the finite volume of a computational cell used to resolve the hydrodynamic flow. Due to this, each computational cell contains at least one active particle at a given time that is distinct from the approaches that resolve the particles using a finite number of computational cells like the one presented \cite{utkin2025particle}. Without a doubt, such an approach could provide additional details about the process of each particle interaction with a surrounding fluid, but here we accept simpler model as a first approximation.

\subsection{Numerical method}
\label{s_sec22}

The incompressible flow equations  (\ref{eq:4},\ref{eq:5}) are solved using the numerical method proposed in \cite{McGrattan} for low-Mach flows. Using the vector identity $\left( {\vec u \cdot \nabla } \right)\vec u = \nabla {\left| {\vec u} \right|^2} / 2 - \left( {\vec u \times \vec \omega } \right)$ (here $\vec{\omega}=\nabla\times\vec u$) and defining the stagnation energy per unit mass as $H \equiv {\left| {\vec u} \right|^2}/2 - p/\rho$ the momentum balance equation can be written as:

\begin{equation}
\frac{{\partial \vec u}}{{\partial t}} + \vec F + \nabla H = 0
  \label{eq:6}
\end{equation}

where 
\begin{equation}
\vec F =  - p\nabla \left( {\frac{1}{\rho }} \right) - \frac{\mu }{\rho }{\nabla ^2}\vec u + \frac{1}{{\rho V}}\sum\limits_{{N_p}} {{m_p}} \frac{d{\vec u_p}}{dt}
\label{eq:7}
\end{equation}

Finally, the Poisson equation for $H$ is derived: 

\begin{equation}
{\nabla ^2}H =  - \nabla  \cdot \vec F
\label{eq:8}
\end{equation}

The numerical solution for the liquid velocity $\vec u$ and pressure $p$ is obtained using a predictor-corrector algorithm described in \cite{McGrattan}, based on Equations (\ref{eq:4}, \ref{eq:5}, and \ref{eq:8}). The Poisson equation (\ref{eq:8}) is solved with a cell-centered multigrid method \cite{Briggs, mohr2004cvs}. 

The governing equations for the Lagrangian particles (Eqs.~\ref{eq:1} and \ref{eq:2}) are solved using the following procedure. First, the linearized form of the particle momentum equation is solved:

\begin{equation}
\vec u_p^{n + 1} = \vec u + \left( {\vec u_p^n - \vec u - {\tau _{St}}\left( {{\vec F}_{p}} \right)} \right){e^{ - \frac{{\delta {t_p}}}{{{\tau _{St}}}}}} + {\tau _{St}}\left( {{{\vec F}_{p}}} \right)
\label{eq:9}
\end{equation}

\noindent where $\delta t_p$ is the time step for particle integration. The particle coordinates are updated at each time step as follows:

\begin{equation}
\vec x_p^{n + 1} = \vec x_p^n + \frac{{\delta {t_p}}}{2}\left( {\vec u_p^{n + 1} + \vec u_p^n} \right)
\label{eq:10}
\end{equation}

The algorithms described for solving the governing equations of fluid dynamics and for Lagrangian particle tracking are implemented in the software package NRG \cite{NRG}, developed by the authors.

\subsection{Resolving time scales}
\label{s_sec23}
Within the framework of the considered problem, the dynamics of the liquid phase occurs to be less intense than the dynamics of the particles. This results in a certain difference in computational time scales for the continuum and particulate phases, making the adequate determination of the relationship between the modeling time steps crucial. Here we analyze an approach in which the liquid phase is considered frozen while the particles move. During the time step $\delta t_f$ used to advance the liquid phase dynamics, the following procedure for the Lagrangian dynamics of each individual particle $p$ is implemented:

\begin{itemize}

\item Calculate the particle time step $\delta t_p = \delta t_f / \tau$.
\item {For $\tau$ substeps:}

\begin{itemize}
\item Calculate forces acting on each Lagrangian particle:
    \begin{itemize}
    \item Calculate the active force components:
    
    {${F}_{active, x} = A_{active} \cdot \cos\left( \theta_p \right) / \delta t_p^{0.5}$,
    
    ${{F}_{active, y} = A_{active} \cdot \sin\left( \theta_p \right))} / \delta t_p^{0.5} $}, 

    where $A_{active}$ is the active force magnitude, $\theta_p$ is the angle defining the direction of the active force action which varies each time substep by $\eta_{noise} \cdot \pi \left( 2\phi - 1\right)$ that models the rotation of each particle, $\phi \in [0,1]$ is a uniformly distributed pseudo-random number. 
    \item Calculate $\vec{F}_{\text{col}}$ using an elastic collision model for spherical particles when trajectories intersect within $\delta t_p$.
    
    \end{itemize}

\item{Advance particle velocities and coordinates using Eqs.~(\ref{eq:9}) and (\ref{eq:10}) over the substep $\delta t_p$. The liquid phase is considered frozen during each particle substep.}

\item{Accumulate the body force $\vec{f_b} = \frac{1}{{\rho V}}\sum\limits_{{N_p}} {{m_p}} \frac{d{\vec u_p}}{dt}$  in each computational cell of the Eulerian grid for the continuum phase.}

\end{itemize}

\item{The resultant body force is incorporated into the right-hand side of the liquid momentum equation (\ref{eq:5}) to advance the liquid phase parameters using the procedure described in Subsection \ref{s_sec22}.}
\end{itemize}

\begin{figure}[ht!]
	\centering\includegraphics[width=0.75\linewidth]{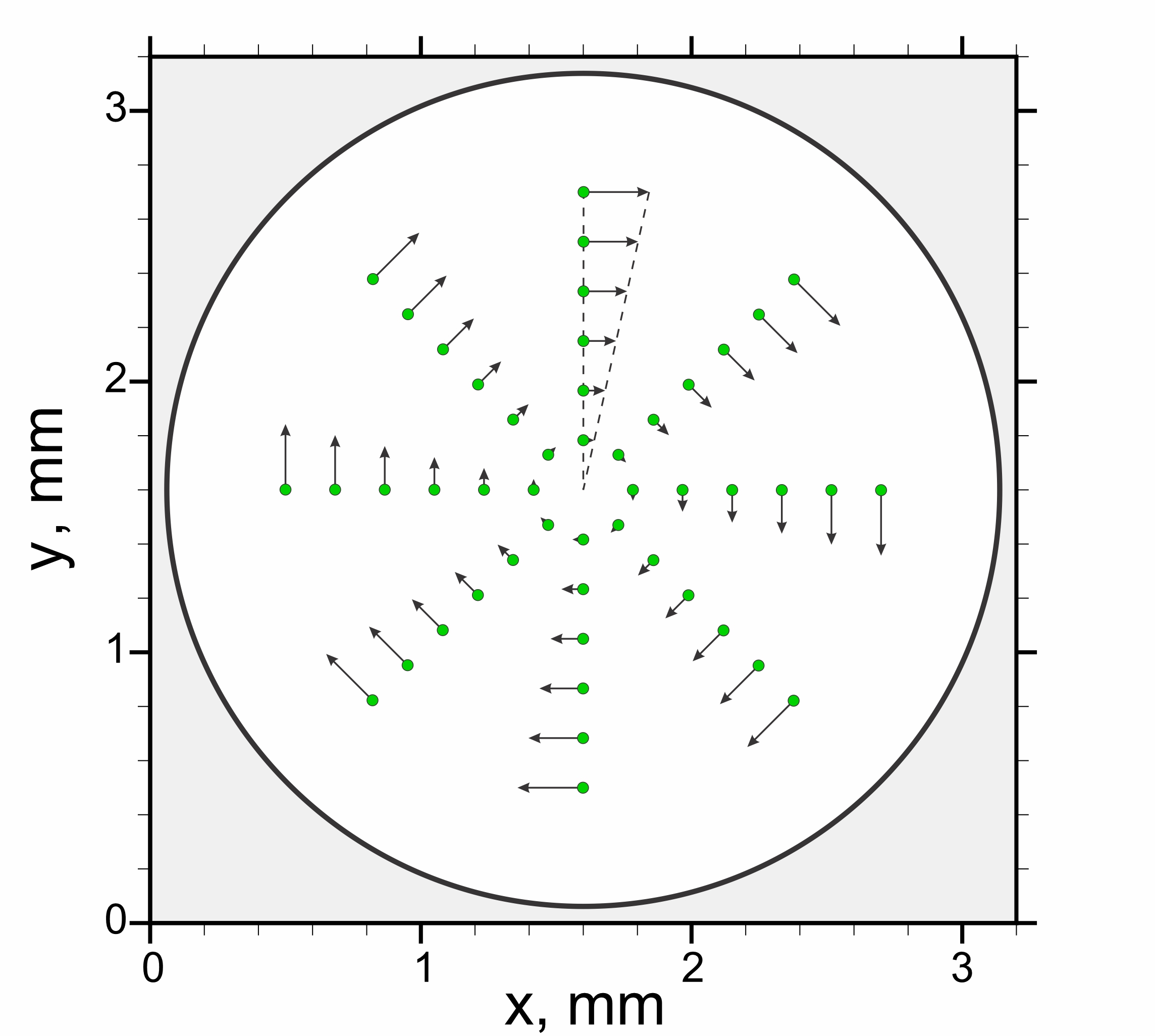}
	\caption{\label{figure01} Schematic of the test problem. The green dots indicate the particles' initial positions (particles diameter $d_p=40$~$\mu$m), and the arrows represent the particles' velocity vectors.}
\end{figure}

\begin{figure}[ht!]
	\centering\includegraphics[width=0.95\linewidth]{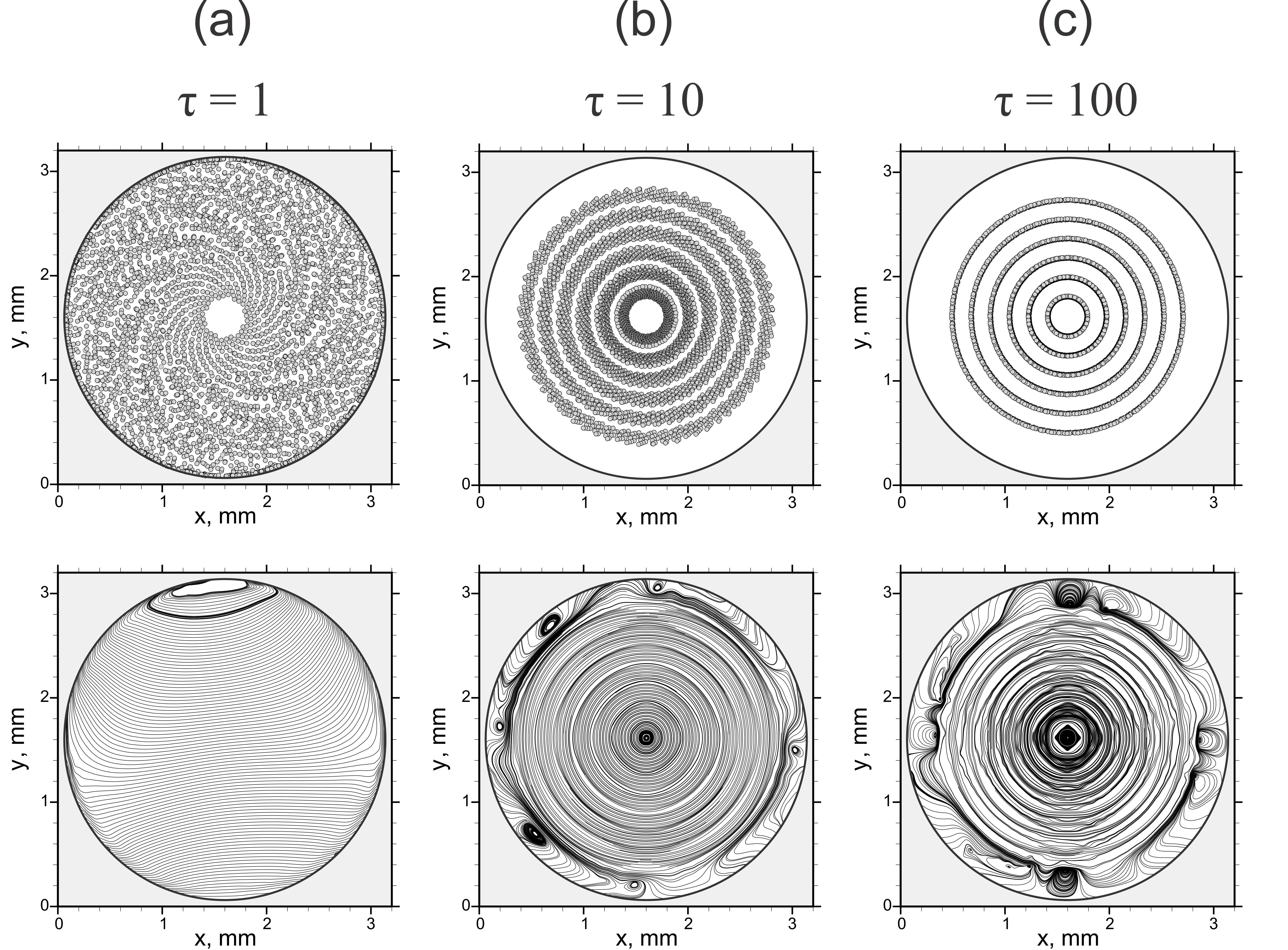}
	\caption{\label{figure02} Results of the test problem simulation. Top row: particle positions at time instances from 0 to 1~s, $\Delta t=10$~ms. Bottom row: fluid streamlines at t = 1~s. Columns correspond to (a) $\tau=1$, (b) $\tau=10$, and (c) $\tau=100$.}
\end{figure}

To illustrate the effect of time scale resolution, a test problem is proposed. In this setup, the liquid is confined within a circular domain of diameter 6.4~mm, with particles uniformly distributed as shown in Figure~\ref{figure01}. The particle velocity is prescribed to be linearly dependent on the distance from the domain center $(x_0, y_0)$ according to $\vec{u}_p = 100.0\cdot(y-y_0, -(x-x_0))$. Figure~\ref{figure02} presents the results for varying numbers of particle substeps $\tau$.

When the particle time step equals the liquid phase time step ($\tau=1$, Fig.~\ref{figure02}a), the particles cannot maintain the prescribed circular trajectories. Due to insufficient temporal resolution of velocity corrections, the particles deviate from their closed paths and propagate in spiral patterns, eventually accumulating near the domain boundary. This results in the absence of pronounced vortical motion in the carrier fluid phase. Increasing the number of substeps enables particles to more closely follow their prescribed trajectories, facilitating the development of vortical flow in the liquid. For $\tau=100$, the particles maintain circular trajectories, and their collective interaction with the fluid becomes sufficiently focused to generate a well-defined vortical flow in the carrier phase.

This test mimics a scenario with a predetermined active force that continuously corrects particle velocity. The results demonstrate that due to the discrete nature of the numerical algorithms employed for modeling two-phase active media, the resolution of time scales critically influences both the emerging flow patterns and the dynamic modes of the active colloid motion. 

\subsection{Problem setup}
\label{s_sec24}
The dynamics of an active colloid (suspension) are studied in a two-dimensional approximation. On the one hand, two-dimensional active systems are of certain interest themselves \cite{YANG201733}. On the other hand, the two-dimensional formulation can be treated as a first approximation of active colloid flow in a thin layer neglecting the interactions between the suspension and the substrate. While such interactions would introduce additional dissipation of kinetic energy within the thin liquid layer, they are not expected to qualitatively alter the system behavior within the scope of this model. The computational domain is a square region with a characteristic size of 3.2~mm. Initially, particles of specified size ($d_p$), mass ($m_p$), and volume concentration ($n_p$) are uniformly distributed throughout the domain. Preliminary calculations were performed in larger domains (e.g., 12.8~mm $\times$ 12.8~mm \cite{KivYak25}), which revealed no significant differences in the system's characteristic behavior across spatial scales. Consequently, the smaller domain was selected to reduce computational expense during parameter studies.

A preliminary analysis of the proposed model has been recently presented in \cite{KivYak25}. That study demonstrates that the dynamic behavior of active particles in a suspension is influenced by several key parameters: the particle velocity $u_p$, governed by the amplitude of the active force; the volume concentration $n_p$, which determines the frequency of collisions and the strength of hydrodynamic interactions; the liquid viscosity $\mu$; and the particle size $d_p$ and mass $m_p$, which collectively with the viscosity set the momentum exchange between phases through the Stokes time scale $\tau_{St}$. The liquid density $\rho$ also plays a role, as it together with the viscosity governs the fluid dynamic response to the particle motion. 

The particle diameter is set to $d_p = 40~\mu\text{m}$ as a reference parameter, with magnetite chosen as the material, as it was used in \cite{kononov2024active} to drive active motion under laser radiation. Water is employed as the liquid phase, with reference values for density and viscosity set to $\rho = 1~\text{g/cm}^3$ and $\mu = 9 \cdot 10^{-4}~\text{Pa} \cdot \text{s}$, respectively. The characteristic velocity of the active particles is approximately 1.0~mm/s, which is of the same order of magnitude as velocities observed in experimental systems \cite{KichatovVortex,KichatovSuperfast}; this velocity is varied from 0.25 to 5.0~mm/s in the study. Initially, the direction of the active force acting on each particle is randomly oriented. The volume fraction of particles, $n_p$, is varied from 0.2 to 0.6, corresponding to a change in the total number of particles in the 3.2~mm $\times$ 3.2~mm domain from $N_p = 1600$ to 4900. Prior to the main series of calculations, selected simulations were repeated for larger domains of sizes 6.4~mm $\times$ 6.4~mm and 12.8~mm $\times$ 12.8~mm, with the same volume fractions $n_p = 0.2$ and 0.6, resulting in a proportionally larger number of particles. Analysis of the flow patterns and energy spectra indicates that the results are independent of the domain size. Therefore, the results from the main series of calculations for the smallest domain are presented hereafter.

\section{Results and discussion}
\label{sec3}
As noted above, the parameter $\tau$ in the computational model governs the relative importance of two characteristic time scales: the intrinsic active motion of each particle and the particle's interaction with the liquid phase via viscous Stokes friction. The case $\tau=1$ corresponds to the equality of these two time scales, meaning a particle changes its direction (due to the rotation modeled by the active force) on the same time scale as it transfers kinetic energy to the surrounding liquid. When $\tau=m>1$, the particle can change its direction $m$ times during the characteristic momentum transfer time to the liquid. Furthermore, $\tau>1$ enables the resolution of particle collisions against the background of their motion through the viscous liquid. Therefore, the case $\tau>1$ represents a more realistic model, as it captures the interactions between the swarm of active particles and the liquid phase of the suspension.

\begin{figure}[ht]
	\centering\includegraphics[width=0.75\linewidth]{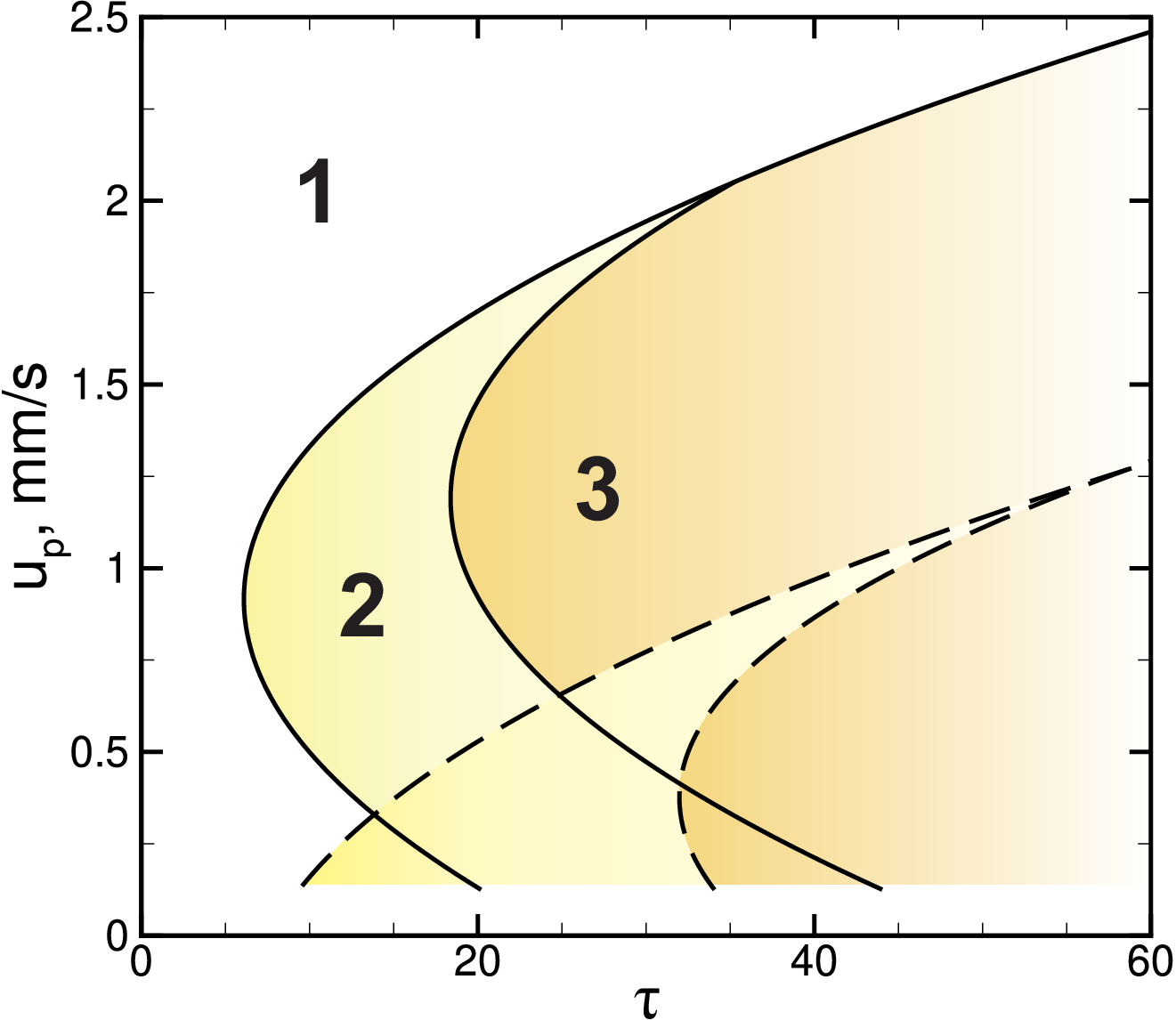}
	\caption{\label{figure1} Diagram of dynamic modes in terms of microswimmers velocity ($u_p$) and parameter $\tau$ for $n_p=$~0.2 (solid lines) and $n_p=$~0.6 (dashed lines). 1 -- Brownian-like motion, 2 -- transient mode, 3 -- collective motion.}
\end{figure}

To better understand the influence of the parameter $\tau$, a series of calculations was performed with varying values of $\tau$, $u_p$, and $n_p$. Figure~\ref{figure1} presents a phase diagram of the active suspension dynamics as a function of these parameters. Three distinct modes are identified: (1) Brownian-like motion of particles against a background of nearly stagnant liquid; (2) a transient mode where collective motion initially emerges but subsequently transitions to a Brownian-like state; and (3) a stable collective motion mode. As the diagram shows, the proposed two-phase model cannot reproduce collective motion at low values of $\tau$, indicating that resolving both physical time scales is crucial for modeling collective effects in active colloids. 

For instance, using $\tau=10$ allows the model to reproduce intrinsic features of active colloids, such as phase transitions occurring at $u_p=0.5$ mm/s and $u_p=1.3$ mm/s. Phase transitions with increasing microswimmer velocity are commonly observed; for example, in systems of bacterial swimmers \cite{Peng2021}, bacterial turbulence emerges when the velocity exceeds a threshold. A similar transition is seen in active emulsions \cite{KichatovVortex}, where ordered collective motion appears only beyond a critical droplet velocity. This phase transition has also been captured by one-phase hydrodynamic models of active matter \cite{linkmann2019phase}, where negative viscosity was used to inject kinetic energy into the liquid phase from subgrid scales representing particle motion. 

The reverse transition, from ordered collective motion back to a Brownian-like state, has been observed in systems with noise-induced rotation \cite{ginelli2010large}. In that study, an increase in noise amplitude led to a loss of order. In the present model, increased noise corresponds to either a higher particle velocity or a decrease in $\tau$ toward unity. Both of these limits, akin to the increased noise in \cite{ginelli2010large}, lead to a transition to a ``run-and-tumble'' mode. At high velocities, particles undergo long runs between collisions, which randomize their directions. Similarly, at $\tau = 1$, particle reorientation occurs discretely at each hydrodynamic time step.

Let us now consider in details all the distinguished modes.

\subsection{Brownian-like mode}
\label{s_sec31}
The Brownian-like mode occurs at both low and high particle velocities (region 1 in Figure~\ref{figure1}). At low velocities, particle motion is dominated by the active force, while collisions occur infrequently due to the low speed. Concerning the hydrodynamics, lower particle velocities result in less momentum transfer to the liquid phase of the suspension. Furthermore, a significant portion of the kinetic energy imparted to the liquid is dissipated through viscous friction. Consequently, the liquid phase motion remains weak and insufficient to affect particle dynamics. Thus, the motion of slow particles is governed solely by the active force, which in the proposed model produces Brownian-like motion characterized by nearly constant speed and random reorientation. 

\begin{figure}[ht!]
	\centering\includegraphics[width=0.95\linewidth]{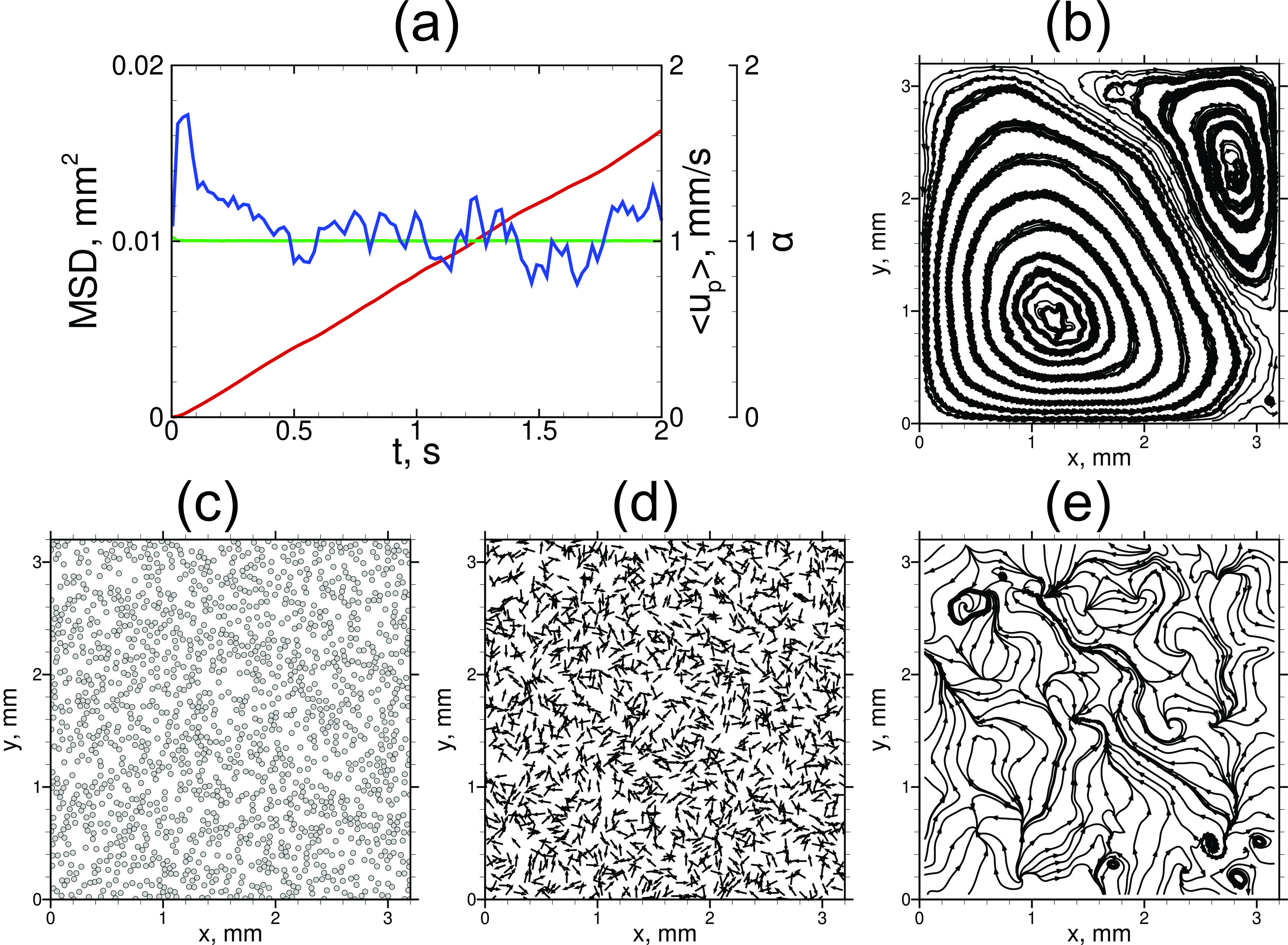}
	\caption{\label{figure2} Analysis of the flow patterns developed in Mode 1 ($u_p=1$ mm/s, $n_p=0.2$, $\tau=5$). (a) Time histories of the mean square displacement of particles ($MSD$, red line), the average particle velocity ($u_p$, green line), and the exponent $\alpha$ characterizing the scaling $MSD \propto t^{\alpha}$ (blue line). (b) Streamlines of the fluid phase flow. (c) Instantaneous spatial distribution of particles. (d) Velocity vectors of the particles. (e) Stream traces illustrating the collective motion of the particles. Frames (b)--(e) correspond to time $t = 2$~s.}
\end{figure}

The mean square displacement ($MSD$) in this mode scales linearly with time ($MSD\propto t^{\alpha}$, $\alpha\approx1$, Figure~\ref{figure2}~$(a)$), while the particles move randomly within the domain  (Figure~\ref{figure2}~$(c),~(d),~(e)$). Figure~\ref{figure2}~$(c)$ shows the instantaneous spatial distribution of particles at $t=2$~s, and Figure~\ref{figure2}~$(d)$ displays their velocity vectors, which exhibit no preferred direction. Streamlines of the velocity field (Figure~\ref{figure2}~$(e)$) indicate that although local structures may form, no collective motion of the entire particle swarm is observed. Concurrently, coherent motion develops in the liquid phase (Figure~\ref{figure2}~$(b)$), but the liquid velocity remains several orders of magnitude lower than the particle velocity, resulting in negligible feedback on the particle dynamics. In this regime, the surrounding liquid acts essentially as a quiescent medium that absorbs particle momentum, converting it to heat through viscous dissipation while sustaining only weak coherent motion.

\begin{figure}[ht!]
	\centering\includegraphics[width=0.75\linewidth]{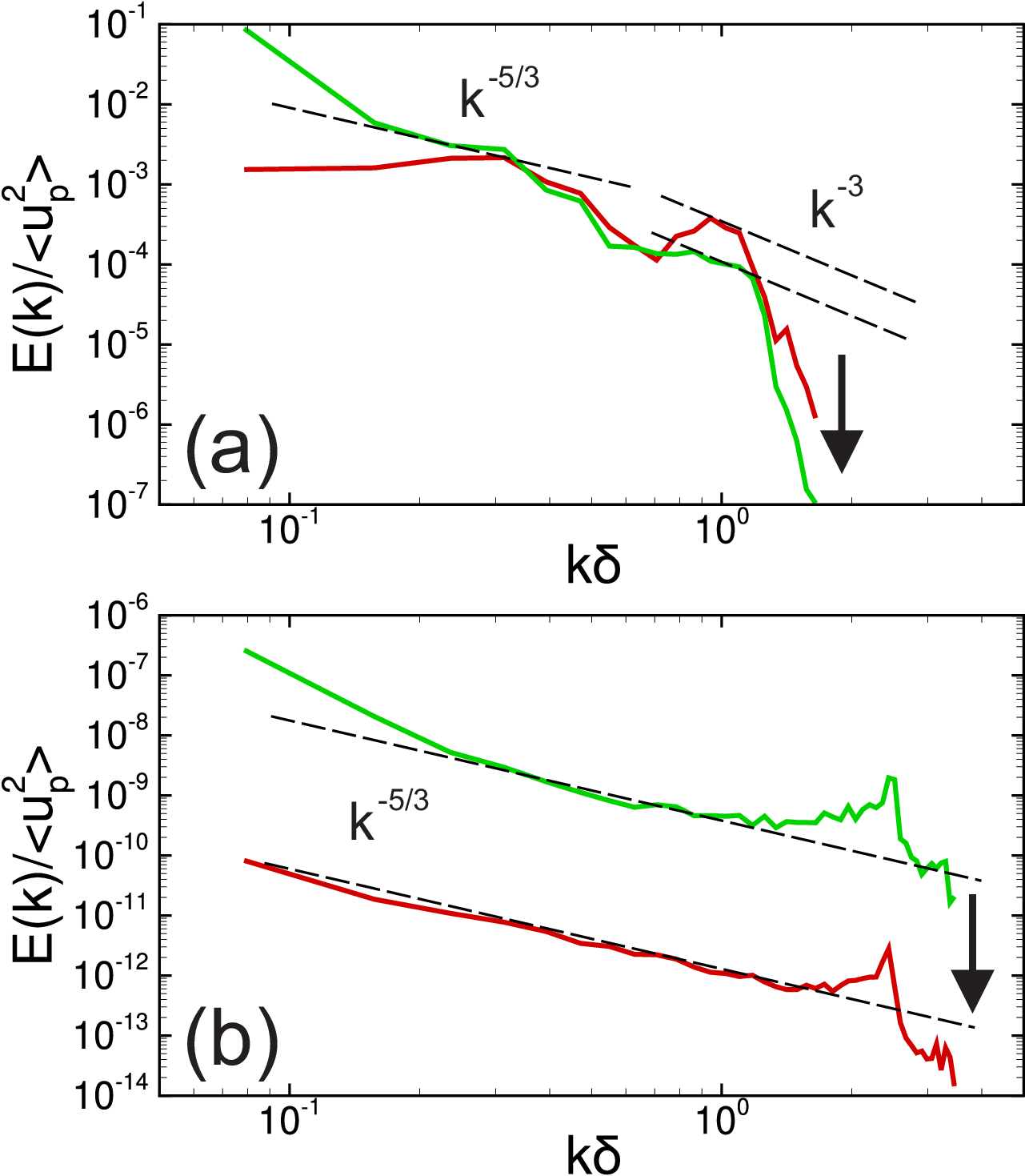}
	\caption{\label{figure3} Energy spectra for the particulate phase (a) and the fluid phase (b). The red curves correspond to Mode 1 ($u_p=1$ mm/s, $n_p=0.2$, $\tau=5$), and the green curves to Mode 3 ($u_p=1$ mm/s, $n_p=0.2$, $\tau=50$). The dashed lines indicate the reference slopes $E(k) \propto k^{-5/3}$ and $E(k) \propto k^{-3}$. Arrows mark the dissipation ranges. Here, $k$ is the wavenumber and $\delta$ is the mean path of the particles.}
\end{figure}

The energy transfer in the system of Brownian motors suspended in the liquid phase is illustrated by the energy spectra of both phases (Figure~\ref{figure3}, red lines). The kinetic energy spectrum of the particles is excited at $k\delta \sim 1$, indicating that the smallest scale of particle motion corresponds to the average interparticle distance $\delta$. In the range around $k\delta \sim 1$, the energy cascade exhibits a scaling behavior characteristic of the inertial range in two-dimensional turbulence, $E(k) \propto k^{-3}$, \cite{kraichnan1967inertial}. At smaller scales ($k\delta > 1$), where particle interactions with the surrounding liquid dominate, the spectrum steepens, which is typical for the dissipation range in developed turbulence. This energy transfer from the suspended particles to the liquid represents a specific dissipation mechanism for the kinetic energy of the particulate phase. The particulate kinetic energy decreases significantly at $k\delta \sim 2\ldots3$, while the liquid phase spectrum shows a local peak at these scales (Figure~\ref{figure3}b).

The kinetic energy spectra of both the particles and the liquid exhibit an inverse energy cascade, characterized by $E(k)\propto k^{-5/3}$, indicating energy transfer from the forcing scales ($k\delta\sim1$ for the particles and $k\delta\sim2\ldots3$  for the liquid) to larger scales (smaller $k$). The most powerful structures in the liquid phase are of the domain size ($L$, $k\sim2\pi/L$), as clearly shown in Figure~\ref{figure2}e. In contrast, the largest structures in the collective motion of the particles are limited to smaller sizes of approximately $0.25L$ ($k\delta\sim0.3$), where a secondary local peak appears in the energy spectrum (Figure~\ref{figure3}a, red line). Consequently, no stable large-scale coherent structures form in the microswimmer motion (Figure~\ref{figure2}d-e). This absence can be attributed to the relatively small amount of kinetic energy transferred to the liquid phase. Furthermore, as this energy is injected at small scales ($k\delta\sim2 \ldots 3$), it is almost entirely dissipated before contributing to larger scales, resulting in extremely weak large-scale flow. Under these conditions, the hydrodynamic feedback from the liquid phase on the particle motion is negligible.

The same Brownian-like mode (Mode 1) is observed at high particle velocities. In this case, the dynamics are dominated by interparticle collisions, while hydrodynamic interactions have little effect on particle motion due to the high speed and frequent collisions that continually randomize their directions. Such a system of rapidly moving particles is analogous to classical Brownian motion.

\subsection{Collective motion mode}
\label{s_sec32}
\begin{figure}[ht!]
	\centering\includegraphics[width=0.75\linewidth]{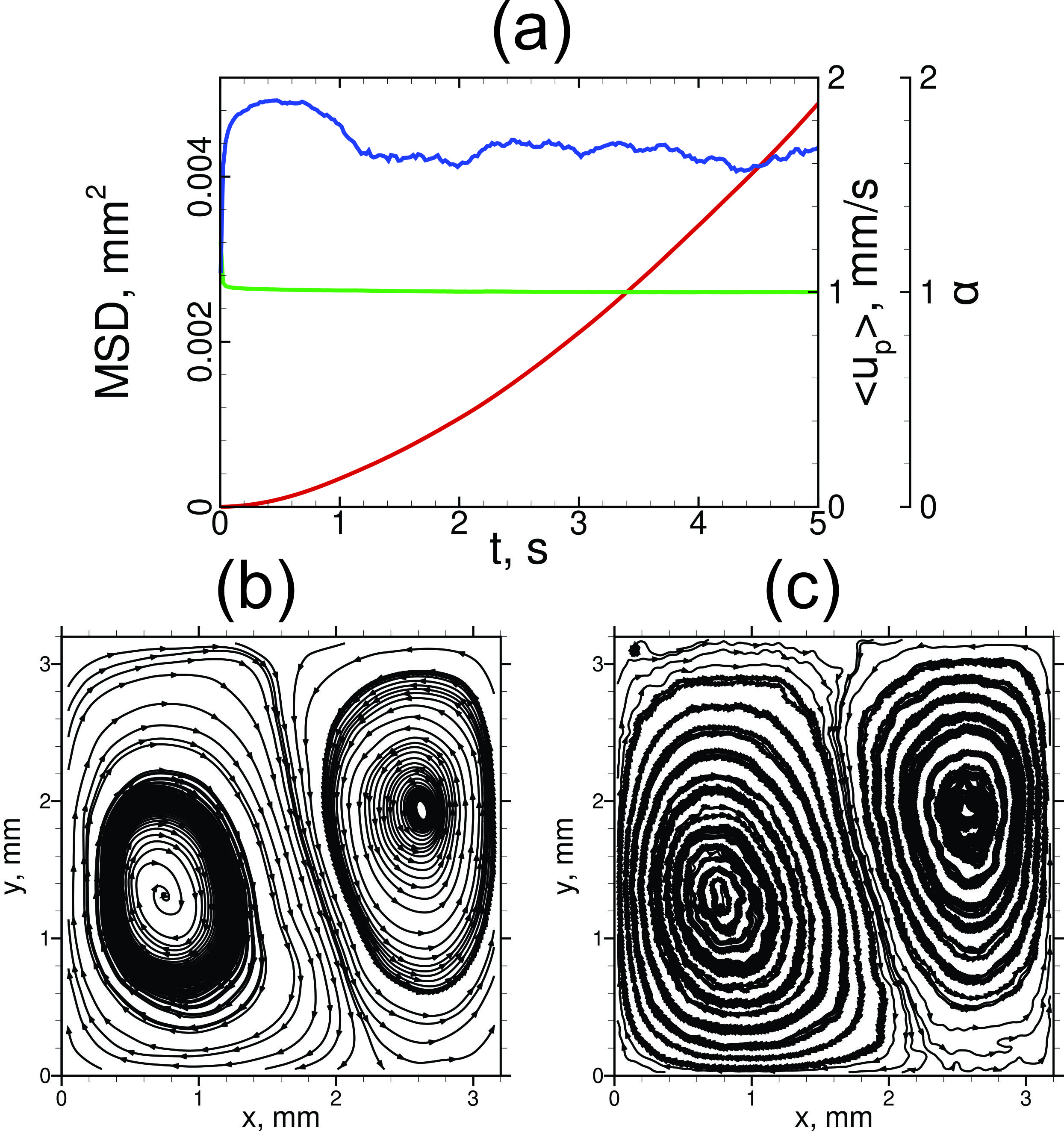}
	\caption{\label{figure_collective} Analysis of the flow patterns developed in Mode 3 ($u_p=1$~mm/s, $n_p=0.2$, $\tau=50$). (a) Time histories of the mean square displacement of particles ($MSD$, red line), the average particle velocity ($u_p$, green line), and the exponent $\alpha$ from $MSD \propto t^{\alpha}$ (blue line). (b) Streamlines illustrating the collective motion of particles. (c) Streamlines of the fluid phase flow. Frames (b) and (c) correspond to time $t = 2$~s.}
\end{figure}

The collective motion mode (Mode 3 in Figure~\ref{figure1}) occurs at moderate particle velocities. In this mode, hydrodynamic interactions play a significant role alongside the active Brownian motion of the particles and their mutual collisions. A coherent collective motion of the particles is observed (Figure~\ref{figure_collective}b), which reflects the flow structures present in the liquid phase (Figure~\ref{figure_collective}c). Similar to Mode 1, the process begins with a Brownian-like motion of active particles that transfer momentum to the surrounding liquid. This energy transfer facilitates the formation of large-scale coherent structures in the liquid via an inverse energy cascade. The resulting coherent flow then exerts a pronounced influence on the particles. Due to the moderate velocity, the randomizing effect of active force rotations is subdued relative to the organizing effect of the hydrodynamic flow. Consequently, the particles become entrained in a coherent motion that replicates the liquid's flow pattern. This establishes a positive feedback loop: the moving microswimmers generate flow in the surrounding liquid, which in turn organizes the microswimmers into coordinated motion.

The collective motion mode exhibits superdiffusive behavior, characterized by a mean square displacement scaling as $MSD\propto~t^{\alpha}$ with $\alpha>$1 (Figure~\ref{figure_collective}a). The fundamental mechanisms of energy transfer between the suspension phases and the energy distribution across spatial scales are analogous to those in Mode 1 (Figure~\ref{figure3}, green lines).  However, the amount of kinetic energy transferred from the active particles to the liquid phase is substantially larger in this mode. This enhanced energy input fosters the development of more pronounced coherent structures, resulting in the formation of stronger vortices that effectively entrain particles into organized motion. It is through this intensified energy transfer that hydrodynamic interactions become significant in governing the collective dynamics of the active particles. Analysis of the energy spectra for both subsystems (Figure~\ref{figure3}, green lines) reveals a coordinated increase in energy following a $k^{-5/3}$ scaling for $k\delta<0.2 \ldots 0.3$, demonstrating the strong hydrodynamic coupling in this wavenumber range.

\subsection{Transient mode}
\label{s_sec33}
It is important to note that even when Mode 1 is realized, a brief superdiffusive regime of active particle motion can be observed (Figure~\ref{figure2}a). In the simulations, this transient behavior is associated with initial interactions between particles, arising from both hydrodynamic coupling and collisions. During this early stage, the forcing scale at $k\delta\sim1$ emerges (Figure~\ref{figure3}a). However, as noted previously, the energy input to the two-phase system in this mode is insufficient to generate substantial fluid motion that could significantly affect the dynamics of the colliding microswimmers. Consequently, the system settles into a sustained Brownian-like motion mode.

\begin{figure}[ht!]
	\centering\includegraphics[width=0.75\linewidth]{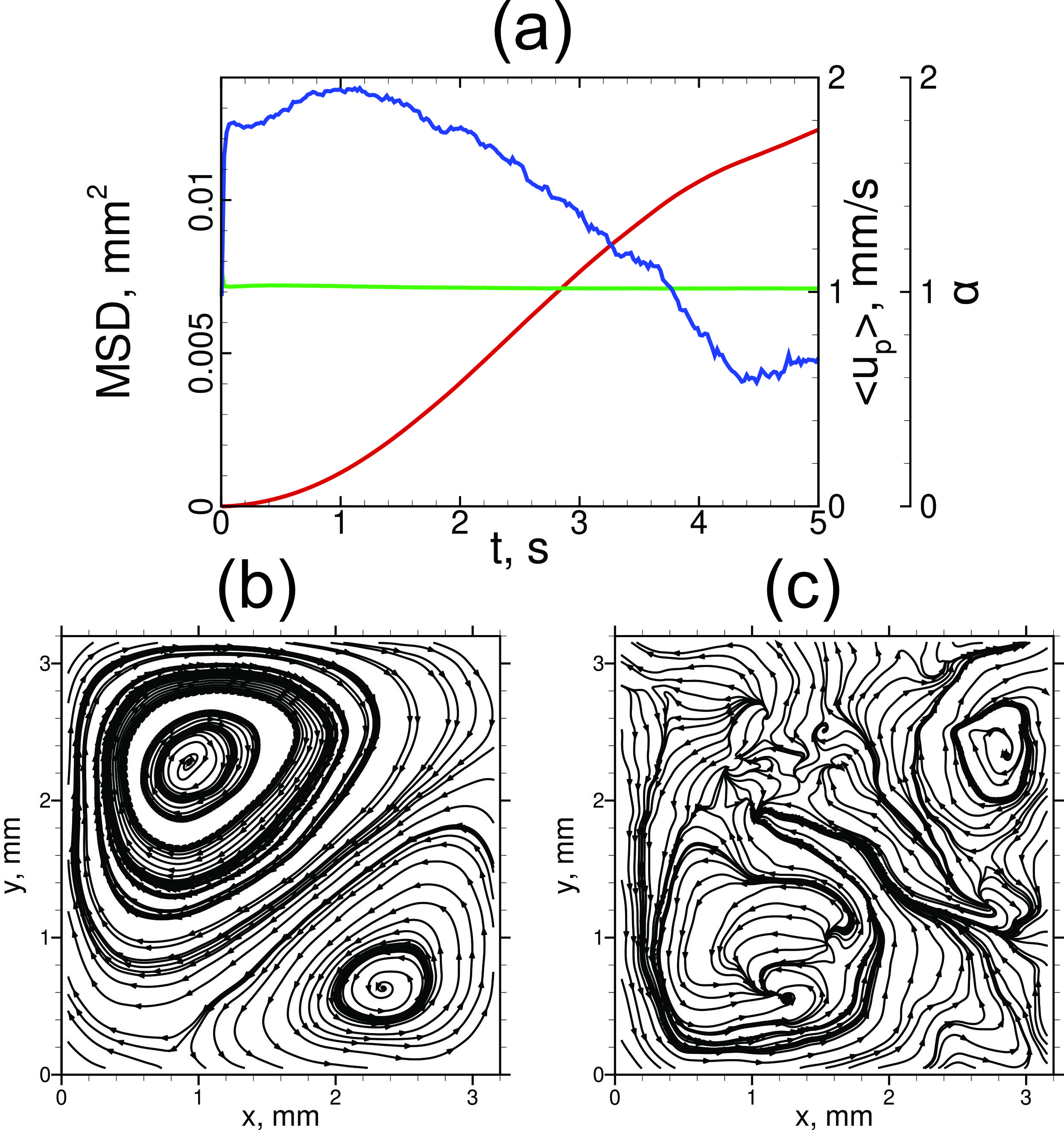}
	\caption{\label{figure_transient} Analysis of the flow patterns developed in Mode 3 ($u_p=1$~mm/s, $n_p=0.2$, $\tau=25$). (a) Time histories of the mean square displacement of particles ($MSD$, red line), the average particle velocity ($u_p$, green line), and the exponent $\alpha$ from $MSD \propto t^{\alpha}$ (blue line). (b), (c) Streamlines illustrating the collective motion of particles at (b) $t = 1$~s and (c) $t = 6$~s.}
\end{figure}

\begin{figure}[ht!]
	\centering\includegraphics[width=0.75\linewidth]{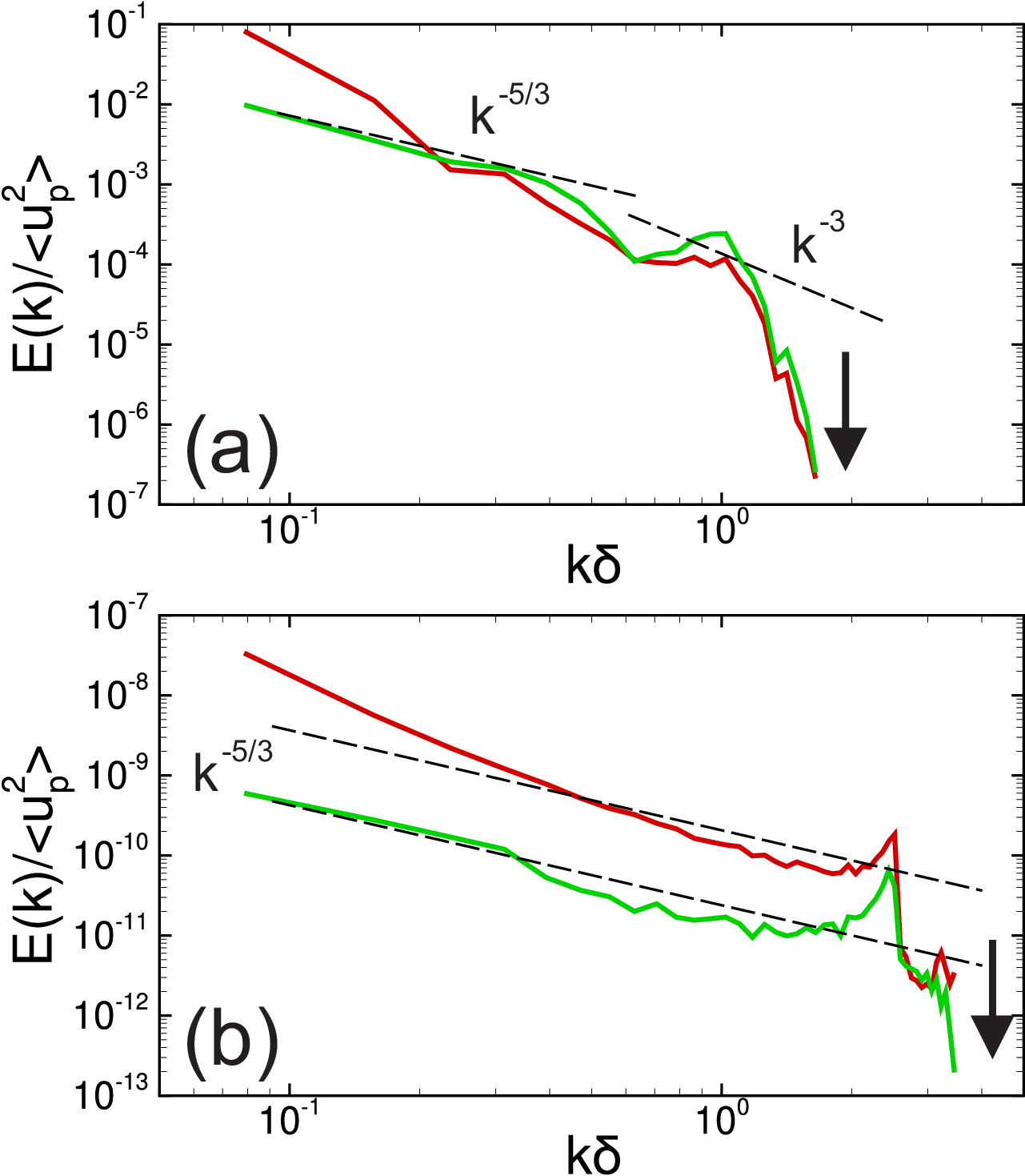}
	\caption{\label{figure4} Evolution of the energy spectra for the particulate phase (a) and the fluid phase (b) in Mode 2 ($u_p=1$ mm/s, $n_p=0.2$, $\tau=20$). Green curves correspond to $t=1$ s, red curves to $t=6$ s. The dashed lines indicate the reference slopes $E(k) \propto k^{-5/3}$ and $E(k) \propto k^{-3}$. Arrows mark the dissipation ranges. Here, $k$ is the wavenumber and $\delta$ is the mean path of the particles.}
\end{figure}

However, the collective motion of the two-phase active medium can also be transient. We refer to this as the transient mode (Mode 2 in Figure~\ref{figure1}). As shown in Figure~\ref{figure_transient}, such a system exhibits collective behavior for a certain period before transitioning to a Brownian-like mode at approximately $t \sim 2\ldots3$~s. Initially, coherent structures form similarly to Mode 3, accompanied by an increase in the kinetic energy of both phases at large scales (Figure~\ref{figure4}, red lines), resulting from particle-fluid interactions and the characteristics of two-dimensional flow. In contrast to the stable Mode 3, the randomizing effect of the active force remains significant in Mode 2. The collective motion is limited in time due to the competition between this randomizing mechanism and the organizing hydrodynamic interaction. At sufficiently low velocities, the inherent randomness of particle motion disrupts the coherence, leading to a transition to the Brownian-like mode. A similar scenario occurs at higher velocities, where the competition between hydrodynamic organization and disruptive particle collisions limits the lifetime of collective motion.

At an increased number of microswimmers, corresponding to a smaller mean free path, the boundaries between the dynamic modes shift toward lower particle velocities. On one hand, hydrodynamic interactions become significant at lower velocities due to the greater momentum transferred per unit volume of the liquid phase. This facilitates the emergence of collective motion at lower microswimmer velocities (Figure~\ref{figure1}). On the other hand, the frequency of collisions increases in denser systems, which also shifts the upper velocity limit for collective motion to lower values, promoting a transition back to the Brownian-like regime (Figure~\ref{figure1}). 

\section{Conclusions}
\label{sec4}

This work presents a novel approach for the numerical analysis of dynamic behavior in active colloids containing microswimmers, utilizing a two-phase hydrodynamic model. We demonstrate that by resolving three fundamental mechanisms: the active motion of the microswimmers, their mutual collisions, and their interaction with the surrounding liquid — the model can reproduce both Brownian-like and collective motion modes.

The Brownian-like modes are primarily driven by the stochastic nature of the active motion and by collisions, whereas the collective mode is strongly influenced by the hydrodynamics of the liquid phase. The motion of the microswimmers inevitably induces flow in the liquid, which follows the laws of hydrodynamics. The kinetic energy transferred from the microswimmers to the liquid is converted into large-scale coherent motion. Once the microswimmer velocity exceeds a critical threshold, a positive feedback loop is established: the liquid motion begins to organize the microswimmers, entraining them into coherent structures. At even higher velocities, however, collisions between microswimmers become dominant, limiting the hydrodynamic organization.

The presented numerical results at least qualitatively correlate well with the experimental data on the evolution of active colloids in thin layers with microswimmers self-organization into the coherent motion \cite{KichatovVortex,senoshenko2024}. As well as in the cited papers, proposed two-phase model reproduces the vortex motion of active particles (or droplets) distinguishing the particular role of hydrodynamics. Thus, the two-phase hydrodynamic model proves to be a promising tool for simulating and analyzing active colloidal systems. Crucially, all three fundamental mechanisms must be adequately resolved in time and space to accurately capture the various dynamical modes. In addition a certain set of validation tests should be elaborated based on experimental data for particular active colloid system.

\section*{Acknowledgments}
The work is funded by the Russian Scientific Foundation (Project №~24-12-00345).


\bibliography{bibliography}





\end{document}